\documentclass[11pt]{article}

\input epsf

\usepackage{amsmath}

\newcommand{\qed}{\rule[-0.2ex]{0.3em}{1.4ex}}

\newenvironment{proof}
               {\par\vspace{0.5ex}\noindent{\bf Proof:}\hspace{0.5em}}%
               {\nopagebreak
                \strut\nopagebreak
                \hspace{\fill}\qed\par\medskip\noindent}

\newtheorem{theorem}{Theorem}

\newtheorem{fact}{Fact}
\newtheorem{definition}{Definition}
\newtheorem{corollary}{Corollary}

\newtheorem{observation}{Observation}

\newcommand{\Prob}{{\rm\hbox{I\kern-.15em P}}}

\newcommand{\IR}{{\rm\hbox{I\kern-.15em R}}}
\newcommand{\reals}{{\rm\hbox{I\kern-.15em R}}}
\newcommand{\IN}{{\rm\hbox{I\kern-.15em N}}}
\newcommand{\IZ}{{\sf\hbox{Z\kern-.40em Z}}}

\newcommand{\prob}{\mbox{\sf \hbox{I\kern-.15em P}}}

\newcommand{\BibTeX}{{\rm B\kern-.05em{\sc i\kern-.025em b}\kern-.08em
    T\kern-.1667em\lower.7ex\hbox{E}\kern-.125emX}}

\begin{document}

\title{An Online Algorithm for Generating Fractal Hash Chains\\Applied to Digital Chains of Custody\footnotemark[1]}
\author{
Phillip G. Bradford\footnotemark[2] 
\and 
Daniel A. Ray\footnotemark[2]}

\maketitle

\begin{abstract}
This paper gives an online algorithm for generating Jakobsson's fractal hash chains~\cite{J}.
Our new algorithm compliments Jakobsson's fractal hash chain algorithm for preimage traversal since his algorithm assumes the entire hash chain is precomputed and a particular list of $\lceil \log n \rceil$ hash elements or pebbles are saved.
Our online algorithm for hash chain traversal incrementally generates a hash chain of $n$ hash elements without knowledge of $n$ before it starts.
For any $n$, our algorithm stores only the $\lceil \log n \rceil$ pebbles which are precisely the
inputs for Jakobsson's amortized hash chain preimage traversal algorithm.  
This compact representation is useful to generate, traverse, and store a 
number of large digital hash chains on a small and constrained device.
We also give an application using both Jakobsson's and our new algorithm applied to digital chains of custody
for validating dynamically changing forensics data.
\end{abstract}

\footnotetext[1]{
An extended abstract of this paper is to appear in the {\em Intelligence and Security Informatics 2007 (ISI 2007)} Conference.}

\footnotetext[2]{
Department of Computer Science, 
The University of Alabama, 
101 Houser Hall,
Box 870290, Tuscaloosa, AL 35487-0290. \{ pgb, DanielRay \} @cs.ua.edu }

\paragraph{Keywords:}
Fractal hash chains, hash chain preimage traversal, hash chain traversal,
digital chains of custody, digital forensics, time-stamping.

\section{Introduction}

This paper proposes a digital forensics system for proactively capturing 
time-sensitive digital forensics evidence.
Intuitively, we are trying to build fully digital chains of custody for
digital evidence that provide forensics investigators with opportunities to enhance 
their data's integrity.
Perhaps this system is best applied to monitor suspects that have already been identified.
In some cases, it is important to capture and validate dynamic evidentiary data.
This may range from tracking a very fast virus infection to tracking more slow
changes directly made by human touch.
Even in monitoring slow changes to a file, 
capturing and validating an inappropriate email--as it is generated--is far more convincing 
than just getting a snapshot of it retrospectively.

Generally, there is a web-of-trust for validating classical forensic evidence.
This web may include both witness and expert testimony as well as 
logical decuction given basic facts about a situation.
The evidence in this web is held together by a chain of custody. A chain of custody is careful documentation
of the evidence including details of all transfers of its possession for examination.
A chain of custody is used to authenticate evidentiary exhibits as well as to
verify these exhibits have not been modified.

Our proposed system also initially depends on a web-of-trust. In particular,
we must trust the system administrators or even law enforcement for initiating our system
under proper circumstances. This may be fleshed out in the usual ways using
witnesses, experts, and logical decuction given facts of the case.
We maintain a full hash chain of the dynamic evidence for the following reasons: (1) to document and 
sychronize the dynamic and time-specific nature of change in a system, (2) to allow repeated verifications 
of the data under scrutiny, and
(3) in the cases where the hash elements include diverse swaths of compressed data may give
extensive opportunites for validation using webs-of-trust.

Our system is based on carefully timed hash chains on small constrained tamper-resistant devices.
These timed hash chains are constructed with compressed snapshots of evidence while using 
(suspected) one-way hash functions.
Generally, the amount of data or the length of time for data capture is not known in advance
for proactively monitoring forensics data.
Thus, the constrained nature of the hardware, variable granularity of the data, and unknown: time and number of hash chains
makes it likely that these devices cannot hold sufficient complete hash chains.
Likewise, such devices cannot hold a large set of time-stamps with compressed digital finger-prints
such as hashs.

It is standard forensics practice to use at least three well-known hash functions to verify data integrity in case one or
more of these functions is ever compromised~\cite{Handbook,KH}.
Such integrity checks may be used for verifying a disk's data. In our case, they may also be used to validate 
dynamic changes in a critical file.
Alternatively, hash chains use deferred disclosure to establish time-synchronized authenticity.
A hash chain can chronologically document evidence
by generating a timed hash chain forward while including diverse data of interest in the hash inputs.
Such evidence may be verified by traversing the hash chain backwards while supplying
the correct inputs at the proper times.
This may include data that may be logically deduced about data captured in the broader system itself.
The backwards hash chain traversal allows repeated verification by only going back in the hash chain
as far as appropriate--leaving the other concealed hash elements in the chain for subsequent deferred verification.
Data may be verified by both sides in a trial, in addition to the
forensic investigator.
To preserve data integrity, each time data is verified, each constrained device will release another
hash element generated prior to all already verified hash elements. This deferred disclosure validates
the known elements in the hash chain and it may tie in to different webs-of-trust.

Another approach is to have each device hold the first and last hash element of a hash chain.
Generally, a forensics verifier would start at the beginning of the hash chain and present the appropriate
data to verify the entire hash chain or a large subset of the hash chain.
However, such large subsets of the hash chain should include either the first or last element of the hash
chain for validity and verification.

Assuming the investigator or system administrator that initiated the hash chain-based data capture is trustworthy. The
data then is in the domain of the evidence clerk maintaining the chain of custody.
Subsequently, the hash chain evidence may be called in to question: is this the correct data?
The first challenge to this approach is if the start (or an early element) of the hash chain is discovered by an attacker, then another fictitious
constrained device may be created that `verifies' incorrect or modified data.
In evidence storage, replacing a gun with another with different serial number or bullet grooves is hard not to miss.
That is, given the first element of a trusted hash chain gives an attacker 
an opportunity to generate and fake the rest of the hash chain, except for the last element.
The last hash element may be at the center of the contention.
If there is a single end element of the hash chain that disagrees, whose do we trust?
Thus, we have three hash chains that only reveal their elements on-demand by
deferred validation: the trusted investigator or system administrator can validate the first or early hash elements.
In fact, ideally there would be enough hash elements that
they would not have to reveal the last hash element.
This situation is reminicant of the motivation for Schneier and Kelsey~\cite{SK}'s audit logs.

Another possibility is to determine the first time step that verifies critical data.
However, authenticating the first critical data item requires the prior hash element.
If it is useful to validate this `prior hash element,' we must traverse the hash chain back two hash elements
from the critical data, {\em etc.}
This is the second challenge to storing the first and last hash elements of a hash chain on a constrained device:
repeatedly validating critical data each time the data is examined through deferred disclosure requires us to traverse
the hash chain backwards.

Computer forensics strives to understand the relationships between suspects and events so they can be verified by 
third parties such as jurors in a court of law.
In such situations it is critical to establish strong credence of the validity of digital forensics evidence.
Digital evidence is abstract, ephemeral, time-sensitive, compact, complex, and often encoded.
In some cases, biological evidence has measurable decay characteristics that allow chronologic analysis~\cite{MI}.
Digital evidence does not have such measurable decay characteristics.
Also, digital evidence is easily copied and copies may be readily manipulated to challenge
valid evidence and diminish its credibility.

\begin{figure}
\leavevmode
\epsfysize=2.0in
\centerline{\epsffile{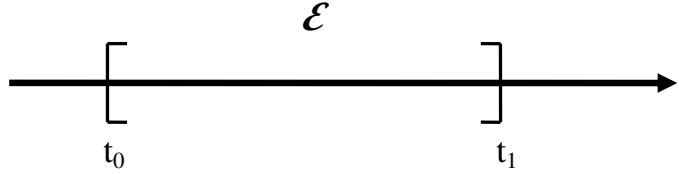}}
\caption{Two Critical Time Boundaries $t_0$ and $t_1$ for Evidence ${\cal E}$. Time increases from left to right and assume $t_1 - t_0$ may be small, for instance as small as $10^{-6}$ of a second.}
\label{Figure0}
\end{figure}

The digital forensics system proposed here is applied to maintaining timed digital evidence.
Figure~\ref{Figure0} illustrates a central application addressed in this paper.
This challenge is the {\em interval time-stamping} problem~\cite{ABSW,HS,Lipmaa,W}.
Consider the left-to-right time line containing the time interval $[t_0, t_1]$.
A central challenge is to demonstrate that the evidence ${\cal E}$ was in a particular user's possession
between times $t_0$ and $t_1$.
The value $t_1 -t_0$ may be very small. 
It is well known that when using public key systems, along with prominent and well distributed data, it is easy to
show ${\cal E}$ existed after time $t_1$.
However, how does one show ${\cal E}$ existed at or before time $t_0$?
One approach is to divide the interval $\left[ t_0 -\frac{t_1 -t_0}{k}, t_1 \right]$ into $k+1$ smaller intervals each of size $\frac{t_1 -t_0}{k}$. 
In each smaller interval $[t_{0,i}, t_{1,i}]$ it may be shown that ${\cal E}$ existed after time $t_{1,i}$, 
for each $i: k > i \geq 1$ where $t_{1,0} = t_{1}$.
Thus, we can show ${\cal E}$ existed before time $t_{1,k} = t_0$.

Throughout this paper, all logs are base~2.

\subsection{Our Contributions}

This paper proposes constrained devices for securing and validating time-sensitive (and dynamic) forensic data.
It assumes tamper-resistant hardware, which may be viewed
as a dependency on a trusted third party.
A special online algorithm is given here to prevent a modify-and-copy attack.
This online algorithm allows computer forensics specialists to maintain the verifiability of 
timed digital evidence.

Take an easy to compute hash function $h$. 
Assume $h$ is one-way~\cite{Goldreich}: So, on average it is intractable to invert $h$.
That is, given $y$ where $y \leftarrow h(x)$ it is on average intractable to find $x$ given
$y$.
Given a hash function $h$ where $x \leftarrow h(y)$, then $y$ is the preimage of $x$.

Next is a $n+1$ element hash chain:
\begin{eqnarray}
v_0\ \stackrel{h(v_1)}{\longleftarrow} \ v_1\ \stackrel{h(v_2)}{\longleftarrow}\ v_2\ \stackrel{h(v_3)}{\longleftarrow} \ \ \cdots \stackrel{h(v_{n-1})}{\longleftarrow} \ v_{n-1}\ \stackrel{h(v_n)}{\longleftarrow}\ v_n.
\label{EQ1}
\end{eqnarray}

Since the hash function is one-way, this hash chain must be initially generated from right to left.

\begin{definition}
Consider a hash chain as described by Equation~\ref{EQ1}.
The value $v_n$ is the seed of the hash chain.

Generating the hash values forward in this order $v_n, v_{n-1}, \cdots, v_0$ is hash
chain traversal. 

Computing hash elements backward in this order
$v_0, v_1, \cdots, v_n$ is preimage traversal. 
\end{definition}

For efficient hash chain traversal see~\cite{HuJaPe2005}.
We always assume the hash function is well known.
Using deferred disclosure of hash elements backwards validates prior knowledge of elements in
the hash chain.
In time step~0, given $v_0$, then waiting to time step~$1$ one can verify that $v_0 = h(v_1)$
indicating with high probability $v_0$ and $v_1$ are from the same source.

At time step $i$, our scheme inputs a chunk of digital evidence ${\cal E}_i$. Let $c({\cal E}_i)$ be the compressed version 
of ${\cal E}_i$. Say $c({\cal E}_i)$ contains a modest number of fixed bits, for instance 160 bits~\cite{Stinson,Vaudenay}.
For example, for the function $c$ we can use a technique such as Merkle-Damg{\aa}rd construction of
a collision-resistant compression function.
Then $v_{i} \oplus c({\cal E}_i)$ is the input to the hash function $h$ giving:

\begin{eqnarray*}
v_{i-1} & \leftarrow & h(v_{i} \oplus c({\cal E}_i) )
\end{eqnarray*}

\noindent
and this process is continued in a carefully timed fashion to give an entire hash chain. The function 
`$\oplus$' may be either xor or concatenation.

If a hash chain is completely exposed an adversary has access to all elements of

\begin{eqnarray*}
C_1 & = & v_n, v_{n-1}, v_{n-2}, \cdots , v_0
\end{eqnarray*}

\noindent
then this hash chain's relative and carefully timed deferred disclosure based authenticity may be easily challenged.
For instance, an attacker may take $v_{n-k}$ and falsify the input ${\cal E}_{n-k}$ by changing it to ${\cal E}_{n-k}'$,
then generate the incorrect hash chain:

\begin{eqnarray*}
C_2 & = & v_n, \cdots, v_{n-k}, v_{n-k-1}', \cdots, v_0'
\end{eqnarray*}

\noindent
where $v_{n-k-i}' \neq v_{n-k-i}$, for all $i: n-k \geq i \geq 0$.

Now, without proofs of identity or authenticity, 
which chain $C_1$ or $C_2$ represents the authenticated data is not clear. 
Of course, proofs of identity or authenticity are only as good as the systems checking them.

Given a hash chain with hash seed $v_n$ and the last hash element $v_0$.
Suppose we only save $v_n$ and $v_0$ on a small constrained device to validate the data from $n$ different times
\begin{eqnarray*}
{\cal E}_{n-1}, \ \cdots, \ {\cal E}_0.
\end{eqnarray*}

If ${\cal E}_i$ is the first piece of critical data in hash element $v_i$, then 
to validate $v_i$, we must know $v_{i+1}$. But, the next time the same hash chain is validated,
we take ${\cal E}_{i+1}$ and validate it with $v_{i+2}$. As these validation steps are repeated each time
investigators weigh the evidence,
we will be performing a (slow and deliberate) preimage traversal of the hash chain.

The hash chain forward traversal algorithm given in this paper extends Jakobsson's work by augmenting his hash chain preimage backward traversal algorithm~\cite{J}.
See also,~\cite{CJ}.
Our algorithm uses hash chains to validate forensics evidence by generating the hash chain elements forward
\begin{eqnarray*}
v_n, v_{n-1}, \cdots, v_0
\end{eqnarray*}

\noindent
online but only storing $\lceil \log_2 n \rceil$ hash elements for a hash chain of $n$ total hash elements.
The preimages of such a hash chain may then be output using Jakobsson's algorithm
and validated by an evidence clerk.
To best ensure the evidence's credence, it is best for the clerk to only compute the backward preimage
traversal as far as necessary. Then, additional deferred disclosures may be performed
as needed to re-authenticate the validity of the exposed hash elements.

A difficulty is in validating a hash chain by backward preimage traversal in memory times computational complexity of less than $O(n)$ where $n$ is the number of hashes required, see for example~\cite{J,CJ,dD,Sella,MP}. 

This paper gives an online algorithm for generating a forward hash chain traversal while always
storing pebbles to be used for backwards preimage traversal by Jakobsson's algorithm.
The online algorithm grows a hash chain as requested to any length $n$, but never
requires storing more than $\lceil \log n \rceil$ hash elements or pebbles, where $n$ hash elements have been generated
so far. This is important on memory constrained devices.
For any $n$, the $\lceil \log n \rceil$ pebbles can be directly plugged into Jakobsson's algorithm
to start backward preimage traversal for verification.

If $n$ is the number of hash elements stored, other recent methods applicable to emitting
hash elements, double the size to $2n$ to 
store a single additional element--the $(n+1)$-st element.
This is not acceptable here for several reasons: (1) our approach depends on precisely timed hash element generation and
generating $n$ more hash elements may cause a simple and constrained device to miss data collection; and (2) the constrained devices may not have the storage to hold $n$ additional hash elements.

\subsection{Previous Work}
\label{PREV_WORK}

Secure audit logs applied to digital forensics were developed by Schneier and Kelsey~\cite{SK}.
This work assumes three machines: a trusted secure server ${\cal T}$,
an small machine who secure state is untrusted ${\cal U}$ for keeping audit logs and 
a sometimes trusted verifier ${\cal V}$.
The audit log machine ${\cal U}$ is a small constrained machine that is only occasionally securely connected
to ${\cal T}$.
A machine ${\cal U}$ is trusted until it is compromised. If it is compromised, then it cannot change its audit log
or read audit elements before the compromise occurred.
This system uses a hash chain to secure the audit logs on the untrusted machines. But, each
untrusted machine deletes all but the last element in the hash chain as the audit log grows. Only ${\cal T}$ has
the seed of hash chain for verification.
Our online algorithm along with Jakobsson's can be applied to variations of Schneier and Kelsey's audit log system.
This would save space and potentially allow numerous hash chains to serve as audit logs.
They close their audit log with a `normalCloseMessage' to prevent an adversary from
extending it illicitly.

To our knowledge, all hash chain preimage traversal work to date assumes either (1) the hash chain is computed in advance, or 
(2) the length of the hash chain is known or a bound on the length is known in advance, or 
(3) given $2^t$ hash elements, the hash chain may be doubled in size to $2^{t+1}$ hash elements when the $(2^{t}+1)$-st element is needed, for $t$ a positive integer.

The first case does not seem directly applicable to our digital forensics scheme.
In the last two of these cases, there may be a good deal of excess unused memory. 

\subsubsection{Time-Stamping}

Time-stamping is a significant area of research. 
Haber and Stornetta~\cite{HS} gave time-stamping schemes using both hash chains and digital signatures.
This is similar to the hash chain scheme used in this paper.
The first method they give is a linking scheme using hash chains to maintain temporal integrity. All of their
schemes depend on a trusted third party. The trusted third party provides a time-stamping service which
applies a hash function and digitally signs it.
Other related schemes and improvements were given in this paper as well as Bayer, Haber, Stornetta~\cite{BHS} and Haber and Stornetta~\cite{HS2}.
Buldas, Laud, Lipmaa and Villemson~\cite{BLLV} focus on `relative temporal authentication' and give
both time-stamping requirements and new algorithms suited to these requirements. 
Their methods require a trusted third party to provide time-stamping services.
Ansper, Buldas,  Saarepera, and Willemson~\cite{ABSW} discuss linkage based protocols (i.e., hash chain based protocols) compared to hash-and-sign time-stamping protocols.
Building on work of Willemson~\cite{W}, Lipmaa~\cite{Lipmaa} gives efficient algorithms
to traverse skewed trees.
Our paper uses basic time-stamping by hash chains. Though the focus is on constrained devices and 
digital forensics.

\subsubsection{Hash Chain Traversal}

Our forward hash chain traversal is based on Jakobsson's backward hash chain preimage traversal work, see also Coppersmith and Jakobsson~\cite{CJ}.
Jakbosson gives an asymptotically optimal algorithm to compute consecutive preimages of hash chains.
His algorithm requires $\lceil \log n \rceil$ storage and $\lceil \log n \rceil$ hash evaluations per
hash element output.
This is assuming $O(n)$ preprocessing was used to build the hash chain of $n$ elements.

Coppersmith and Jakobsson~\cite{CJ} give an algorithm with amortized time-space 
product cost of about $\frac{1}{2} \log^2 n$ per hash chain element.
This is also assuming $O(n)$ preprocessing was used to build the hash chain of $n$ elements.
They also give the following lower bound: 
Computing any element of the hash chain in the worst-case requires a time-space 
trade-off $TS \geq  \frac{\log^2 n}{4}$, where $T$ is the number of invocations of the hash function
and $S$ is the number of stored hash elements.

Sella~\cite{Sella} gives a general solution that applies $k$ hash function evaluations
to generate any element in a hash chain, while storing $(k-1) n^{1/(k-1)}$ total hash elements.
His algorithm initially stores hash elements that are at constant intervals of distance $(k-1) n^{1/(k-1)}$.  
Kim~\cite{Kim03,Kim05} gives algorithms that improve Sella's in saving up to half of the space Sella's algorithms
use while keeping the same parametric space and hash evaluation costs.
This means Kim's algorithm uses at most the same space as Jakobsson's algorithm.

Ben-Amram and Petersen~\cite{B-AP} give an algorithm for backing up in a singly linked list
of length $n$ in $O(n^{\epsilon})$ time, for any $\epsilon > 0$. This requires $O(n)$ pre-processing
of the linked list.

Matias and Porat~\cite{MP} give list and graph traversal data structures that allow efficient backward list traversal and very efficient forward traversal. 
Their `skeleton' data structures allow complete back traversals in $O(k n^{1/k})$ amortized time given
$k$ elements in storage.
Thus, storing $k= O(\log n)$ elements, their algorithm requires $O(\log i)$ amortized element evaluations
for the $i$-th backward traversal, where $i \leq n$.
They bring out several interesting applications beyond hash
chains. If their data structures are built for lists of $n$ elements, then to accommodate $n+1$ elements
they build a new data structure (list synopsis) for $2n+1$ total elements, see subsection 4.3 of the full
version of~\cite{MP}.

\subsection{Structure of this Paper}

In the remainder of the paper we develop the ideas behind our proposed approach. 
Section~\ref{CoC} gives our model.
Section~\ref{JA} reviews Jakobsson's algorithm in detail. We give proofs of correctness
for this algorithm in the appendix. 
In order to prove the correctness of our algorithm we found it necessary to 
supply detailed proofs of Jakobsson's algorithm (which were not given directly in Jakobsson's paper).
Section~\ref{OnLine} introduces the specifics of our online algorithm, discusses how it interfaces with Jakobsson's
amortized algorithm, and gives a proof of correctness.  In Section~\ref{Future} we give some conclusions.

\section {Chains of Custody: Physical and Digital}
\label{CoC}

Next is a formal definition of a chain of custody.

\begin{definition}
{\sf
A {\em chain of custody} is a detailed  account documenting the handling and access to evidence.
}
\label{Classic_CoC}
\end{definition}

We quote Colquitt~\cite[Page 484]{Colquitt} on the purpose of a chain of custody:

\begin{quote}
``The purpose, then, of establishing a chain of custody is to satisfy the court that it is reasonably
probable that the exhibit is authentic and that no one has altered or tampered with the proffered physical 
exhibit.''
\end{quote}

A chain of custody, as described in Definition~\ref{Classic_CoC}, may sometimes be referred to as a {\em classic}
chain of custody.
Maintaining a chain of custody is a standard practice investigators use to inextricably 
link the evidence that ties a crime to the suspects. 

Digital evidence is often stored using a classical chain of custody. 
For example, documenting when a particular individual first picked up a disk drive with critical evidence, 
the state of the disk drive, to whom and when they transferred the drive, {\em etc.}

Digital evidence is extraordinarily easy to copy. Using standard techniques, each copy of digital evidence is easily
authenticated.
Physical evidence is somewhat different~\cite{HS,MI}.
Consider a crime committed by using a gun. Manufacturing a new $.357$ Magnum Ruger Blackhawk Flattop revolver with an identical `look,' serial number, and bullet grooves, to one used in a crime is extraordinary work. Just finding experts to replicate such physical evidence would generally
leave a substantial paper trail. Moreover, biological evidence generally also provides time frames.
Thus, classic chains of custody often focus on the basic identification,
authentication, uniqueness, and time.
Digital data lacks such uniqueness and timing characteristics. That is, many copies may be made of digital evidence
both for legitimate and illicit reasons. Furthermore, timestamps alone may not be sufficiently convincing.

\begin{definition}
{\sf
A {\em digital chain of custody} is the information preserved about the data and its changes that 
shows specific data was in a particular state at a given date and time.  

A {\em DCoC (Digital Chain of Custody)} is a small constrained device for holding, authenticating, and verifying
a digital chain of custody.
}
\label{DCoC}
\end{definition}

Take a one-way hash function $h$. 
Suppose $h$ has deterministic and tightly bounded time complexity.
There are some hash functions with bounded time complexity that are already in use:
the RSA SecurID~\cite{RSA} as well as systems like TESLA~\cite{TESLA,PerrigTygar} depend on
timed hash functions.
This paper goes one step further advocating extremely precisely timed hash functions.

This paper is focused on several aspects of evidence maintenance that are related to time stamping, see for example~\cite{dD}.
One side of the challenge is to show a piece of evidence existed after a particular point in time.
Using the Merkle-Damg{\aa}rd technique, the evidence may be combined with non-predictable, widely distributed, and time sensitive documents such as a newspaper or financial market data\footnote{As financial markets become more automated
and distributed, it may be feasible for the trading volume to be granulated down to minute fractions of a second. For example, in 2004 the average NYSE trading day volume per second was
about $1,500,000,000/(60 \cdot 60 \cdot 6.5)$ or more than 64,000 shares per second. See www.nyse.com. The volume is known down to the individual share. This highly granular data may be widely distributed.}.

Similar discussions to the next definition can be found in several places, for example~\cite{HS,Vaudenay}.

\begin{definition}
{\sf
Data that is non-predictable, widely distributed, verifiably stored, and time sensitive is {\em socially bound} data.
}
\end{definition}

A critical issue is that socially bound and highly granular data are not common.
In certain situations, socially bound data must be highly granular, for example in milliseconds.
Thus, carefully timed hash chains may be used as a proxy for socially bound data. See also, timestamp linkages in~\cite{HS}.

\subsubsection{The Adversary}

The adversary this paper assumes is either (1) an untrustworthy verifier; or (2)
a general attack against known hash functions on a DCoC\@.

Suppose our system computes an $n$ element hash chain $v_n, \cdots, v_0$, and say $n > t$.
An untrustworthy verifier may get an element of a hash chain $v_t$ along with the files of digital evidence 
$E = {\cal E}_{t}, \cdots, {\cal E}_0$.
The adversary may illicitly modify the evidence to $E' = {\cal E}_{t}', \cdots, {\cal E}_0'$
and compute a `competing' hash chain using $v_t$ and $E'$.
Thus, adding doubt to the validity of the real evidence.
Provided $t < n$, and the one-way hash function $h$ is not broken, we can validate
the hash chain based on $E$ and not $E'$ by producing $v_{t+1}$ and showing
that $h(v_{t+1} \oplus c({\cal E}_{t+1})) = v_t$.

The issue of a general breach of a hash function is primarily dealt with by
following the forensics policy of having at least three different known
hash functions for the data. This procedure is to ensure trust in the hash functions in case one of them is no
longer trustworthy~\cite{Handbook}.

\subsection{Overview of Our Approach}

This paper assumes a model consisting of at least three constrained devices (DCoCs or cards) each having a different hash function.
These constrained devices may be interfaced using a USB 2.0, for instance.
In any case, these devices have a small processor as well as limited memory.
They are tamper resistant.

\subsubsection{Data Capture}

The DCoCs may each interface through separate USB 2.0 ports simultaneously to gather common snapshots of
critical system states and memory.
Each DCoC may be physically secured and transported independently by a different
member of law enforcement.
We assume these devices will be mass produced so that each DCoC has unique authentication software.
(They should also have unique physical identification.)
Different members of law enforcement should handle each card making it less likely that all 
cards may be manipulated by a single member of law enforcement. 
Furthermore, from physical identification, the DCoCs may become part of the classical chains of custody.

Finally, each of the DCoCs may periodically authenticate each other while capturing data.
Records of these authentications can be included in the hashed values.
This authentication should be zero-knowledge interactive proofs of identity such as the Feige-Fiat-Shamir protocol~\cite{FFS}. Such a proof of identity is important in this application domain since 
these proofs of identity are not transitive. 
Thus, a fake DCoC cannot impersonate a real one by just mimicking its proof of identity.

\subsubsection{Forensic Data Verification}

The plaintext evidence and diversifying data has been stored as the files $M$ in plain sight on storage such as an optical disk.
Different states or snap-shots of the evidence will be periodically concatenated into timed inputs of the hash chain.
Suppose the hash computations run at a fixed and known speed on tamper resistant hardware.
Thus, computing a hash chain on this special hardware can be used to certify the initial data
states at specified times, see also Haber and Stornetta~\cite{HS}.

The verification procedure is done by an evidence clerk and consists of the following steps:

\begin{enumerate}

\item
Connect all three DCoCs to a secured verification machine. Initially, each DCoC is authenticated to all other DCoCs using 
a zero-knowledge interactive proof of identity.

\item
The DCoCs only output hash elements backwards on an as-needed basis to verify the evidence under consideration.

\item 
As hash chain elements are output, the verification must continue with each hash element step.
If a DCoC cannot authenticate another device, then it will alert the evidence clerk or even
shut down.

\item
The hash elements from a DCoC are sent out from left to right:
\begin{eqnarray*}
h^n(v_n \oplus c({\cal E}_n)), \ h^{n-1}(v_{n-1} \oplus c({\cal E}_{n-1})), \ \cdots, \ h^t(v_{t} \oplus c({\cal E}_{t})),
\end{eqnarray*}
for $n-1 \geq t$. Moreover, $h^n(v_n \oplus c({\cal E}_n))$ is verified by computing 

\begin{eqnarray*}
h(h^{n-1}(v_{n-1} \oplus c({\cal E}_{n-1})) \oplus c({\cal E}_n)),
\end{eqnarray*}
and so on.

\end{enumerate}

\subsubsection{Putting It All Together}

Given the tamper resistant hardware, just storing data with associated time stamps may quickly overflow a constrained
device--especially if the time granularity is very fine.
Thus, compressed representation of hash chains can be used to verify time stamps along with the authenticity of other 
constrained devices.

It is possible to publicly post data from a system under forensic investigation by way of a proxy server.
This data may be signed by a small device and posted to a public (or private and trusted) location.
Small devices may have trouble signing large amounts of data due to their constraints.
Furthermore, for very fine time granularity, fast and consistent network bandwidth 
may not be available.
In some contexts, if data is captured before a formal investigation is initiated, then it may be important 
to keep the suspects unaware that data is being captured as potential evidence.

A DCoC may run its hash chain until a forensic examiner or law enforcement officer carefully checks it into the physical chain of custody while
noting the time in detail. This allows a forensic examiner to determine the time of the evidence by counting the exact number of hashes.
A key motivator of our online algorithm is in proactive forensics the value of $n$ is not known in advance. 
Where $n$ is the number of time-steps recorded by the DCoCs.
In the case
of a digital chain of custody, it can not be known precisely how long it will take to get potential evidence
to an evidence clerk.
Or how long an evidence clerk will take to validate the data, {\em etc.}
This precise time as well as the interval between hashes, computed by our algorithm, will have to be 
known in advance to fake a hash chain. 

The preprocessing phase of Jakobsson's amortized algorithm must know the value of $n$ in advance of when 
it is run~\cite{J}.
Our new algorithm prepares the appropriate $\lceil \log n \rceil$
pebbles for Jakobsson's algorithm, for any $n$. During hash element generation, for any $n$ the online algorithm never
stores more than $\lceil \log n \rceil$ pebbles. Once the online
algorithm has no more requests to generate new hash elements, then the amortized algorithm~\cite{J} 
may be immediately started.

\section{Jakobsson's Algorithm}
\label{JA}
This section reviews Jakobsson's Algorithm~\cite{J}.
The proofs of correctness are in the Appendix. Some of these proofs are used in the results
of Section~\ref{OnLine}.

\subsection{A Review of Jakobsson's Algorithm}

Given a hash chain with 16 elements, Jakobsson's algorithm is initialized as indicated
in Table \ref{tableJ}.  
In this Table, hash element $16$ is the seed of the hash chain. Where the preimages
are exposed sequentially in the following order: $1, 2, \cdots, 16$.
In general, the position of a hash element in a hash chain ranges from $1, \cdots, n$ where hash element $1$ is the first element to be exposed in a pre-image traversal. Thus, hash element $n$ is the seed of the hash chain.

\begin{table*}[here]
\begin{center}
\begin{tabular}{l||c|c|c|c|c|c|c|c|c|c|c|c|c|c|c|c} 
Element Position	& 1 
		& 2 
		& 3
		& 4
		& 5
		& 6
		& 7
		& 8
		& 9
		& 10
		& 11
		& 12
		& 13
		& 14
		& 15
		& 16 \\ \hline
Pebble Placement	        &  
				&  $\times$
				&   
				&  $\times$
				&   
				&   
				&   
				&  $\times$
				&   
				&   
				&   
				&   
				&   
				&   
				&   
	          	&  $\times$ \\ 
\end{tabular}
\\
\vspace{3 mm}
\caption{Initial pebble locations for Jakobsson's algorithm}
\label{tableJ}
\end{center}
\vspace{-5 mm}
\end{table*}

Suppose $n = |H|$ is the number of hash values in the entire hash chain. In
this case, Jakobsson's amortized algorithm only stores $\lceil \log n \rceil$
hash pebbles.
 
Suppose $n = 2^{k}$ for some integer $k$.
Now, Jakobsson's algorithm~\cite{J} is given:

\subsubsection{Jakobsson's Setup.}
	Compute the entire hash chain: $v_0, v_1, \ \cdots, \ v_n$.

	For pebble $p[j]$, where $j: k = \log_2 n \geq j \geq 1$:

	\begin{eqnarray*}
	p[j].{\bf Start\_Increment} 	& \leftarrow & 3 \times 2^{j}\\
	p[j].{\bf Dest\_Increment} 	& \leftarrow & 2^{j+1}\\
	p[j].{\bf Position}  		& \leftarrow & 2^{j}\\
	p[j].{\bf Destination} 		& \leftarrow & 2^{j}\\
	p[j].{\bf Value}  		& \leftarrow & v_{2^j}
	\end{eqnarray*}

	Furthermore, 
	\begin{eqnarray*}
	{\bf Current.Position} 		& \leftarrow & 0\\
	{\bf Current.Value}  		& \leftarrow & v_{0}
	\end{eqnarray*}

\subsubsection{Jakobsson's Main.} This algorithm in Figure~\ref{Fig:JsAlgorithm} updates the pebbles and values.

\begin{figure}[ht]
\begin{center}
\begin{tabbing}
over text\=  foo \= text \= text \= text \kill
\> Given an entire precomputed computed hash chain:\\
\> \> $v_1, v_{2}, \cdots, v_n$, then compute the pebble positions \& auxiliary information.\\
\>1. \> {\bf if} $\mbox{\bf Current.Position} = n$, {\bf then}\\
\>   \> \> {\bf Stop}\\
\>   \> {\bf else}\\
\>   \> \> $\mbox{\bf Current.Position} \leftarrow 1+ \mbox{\bf Current.Position}$\\
\>   \> {\bf endif}\\
\>2. \> {\bf for} $j \leftarrow 1$ {\bf to} $k$ {\bf do}\\
\>   \> \> {\bf if} $p[j].{\bf Position} \neq p[j].{\bf Destination}$ {\bf then}\\
\>   \> 2.1 \> \> $p[j].{\bf Position} \leftarrow p[j].{\bf Position} -2$\\
\>   \> 2.2 \> \> $p[j].{\bf Value} \leftarrow h(h(p[j].{\bf Value}))$\\
\>   \> \> {\bf endif}\\
\>   \> {\bf endfor}\\
\>3. \> {\bf if} {\bf Current.Position} is odd {\bf then}\\
\>   \> \> {\bf output} $h(p[1].{\bf Value})$\\
\>   \> {\bf else}\\
\>   \> \> {\bf output} $p[1].{\bf Value}$\\
\>   \>3.1 \> $p[1].{\bf Position} \ \ \ \ \ \leftarrow p[1].{\bf Position} + p[1].{\bf Start\_Increment}$\\
\>   \>3.2 \> $p[1].{\bf Destination}  \leftarrow p[1].{\bf Destination} + p[1].{\bf Dest\_Increment}$\\
\>   \> \> {\bf if} $p[1].{\bf Destination} > n$ {\bf then}\\
\>   \> \> \> $p[1].{\bf Position} \ \ \ \ \ \leftarrow +\infty$\\
\>   \> \> \> $p[1].{\bf Destination} \leftarrow +\infty$\\
\>   \> \> {\bf else}\\
\>   \> \> \> $p[1].{\bf Value}    \leftarrow {\bf FindValue}$\\
\>   \> \> {\bf endif}\\
\>   \> \> {\bf Sort Pebbles} by {\bf Position}\\
\>   \> {\bf endif}
\end{tabbing}
\caption{Jakobsson's Hash-Chain Pebble Position update}
\label{Fig:JsAlgorithm}
\end{center}
\end{figure}

\begin{figure}
\leavevmode
\epsfysize=2.5in
\centerline{\epsffile{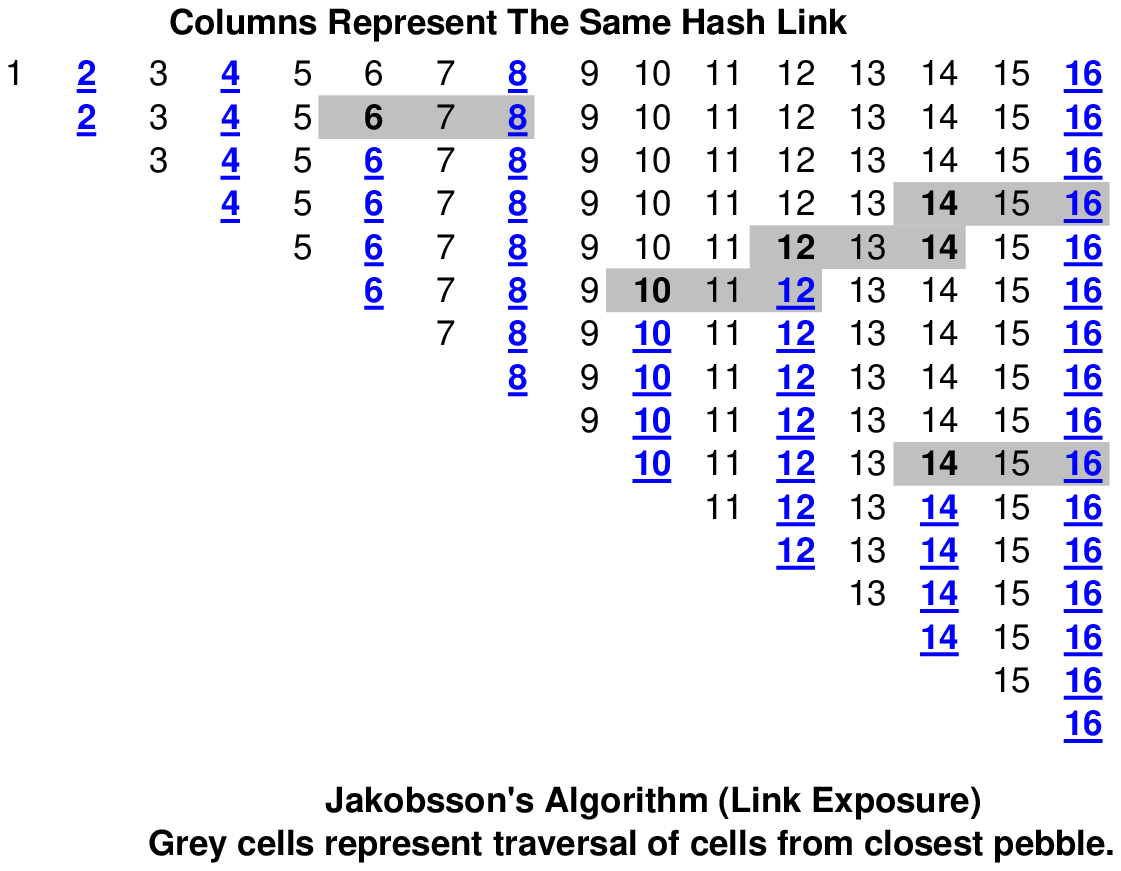}}
\caption{Jakobsson's Algorithm: 
Each row from top to bottom represents the state of a hash chain traversed by Jakobsson's algorithm.  Each row represents
a single hash element exposure. For each row,
bold and underlined positions represent the location of pebbles where {\bf Position = Destination}. Bold positions
represent a temporary location for a pebble in transit only.  When a pebble is reached and moved it is moved
to the back of a grey area and moves to the front of the grey area or sequence of grey areas until it reaches
its destination (and is underlined and in the next row).}
\label{JFigure}
\end{figure}

Using Jakobsson's amortized technique~\cite{J}, a hash chain of size $n$ requires a total of $k$ pebbles where $k = \lceil \log n \rceil$. This amortized algorithm performs $O(\log n)$ hash applications per hash element output. 
Initially, in the setup phase, the pebbles store hash elements from hash-chain positions $2^1, 2^2, 2^3, \ldots, 2^k$, respectively.

\section{Online Output of Jakobsson's Pebbles for any $n$}
\label{OnLine}

Jakobsson's amortized algorithm works to conserve both stored pebbles and hash evaluations.
This allows his algorithm to verify hash values on small sensors. This algorithm assumes
pre-processing where all hash elements are pre-computed, perhaps by a more powerful processor~\cite{J}.

Our aim with the online algorithm is to have a small constrained device that 
generates all requested hash elements. 
The online algorithm broadens the applicability of Jakobsson's amortized algorithm. In particular,
the online algorithm generates all hash elements, but only stores $\lceil \log n \rceil$ pebbles
at any one time. Where $n$ is the total number of hash element generated so far.
Every time a new hash element is generated, no additional hash evaluations must be performed.

These $\lceil \log n \rceil$ hash pebbles are positioned so that at any point the online algorithm is no longer invoked,
then Jakobsson's amortized algorithm can start to be run directly on the pebbles.
Thus, we believe the amortized and online algorithms are complimentary and fit well together.

In Jakobsson's notation each hash element keeps its index throughout the computation,
\begin{eqnarray*}
v_0, \cdots, v_n.
\end{eqnarray*}
This is because $n$ is fixed, so all hash element numbers remain the same.
The pebbles in Jakobsson's algorithm are re-numbered, according to the initial fixed hash element numbers,
and sorted by their positions.

The notation for the online algorithm uses hash element
index notation that changes the index values over time.

For instance, in Jakobsson's preimage traversal algorithm, suppose the initial $n$ hash elements are:
\begin{eqnarray*}
u_1, \ \cdots, \ u_n.
\end{eqnarray*}
Then, after the first element is verified, it is discarded and 
the following hash elements remain:
\begin{eqnarray*}
u_2, \ \cdots, \ u_n.
\end{eqnarray*}

Alternatively, since $n$ is not fixed in our online algorithm, suppose the following list of elements
has already been generated (with only the appropriate pebbles saved),
\begin{eqnarray*}
w_1, \cdots, w_{n}.
\end{eqnarray*}
Then, generating another hash value $w_0 \leftarrow h(w_1)$.
This gives $n+1$ total hash elements, thus for simplicity all hash indices are increased by~1.
This gives the following renumbering,
\begin{eqnarray*}
w_1, \cdots, w_{n+1}.
\end{eqnarray*}

\begin{figure}
\leavevmode
\epsfysize=2.5in
\centerline{\epsffile{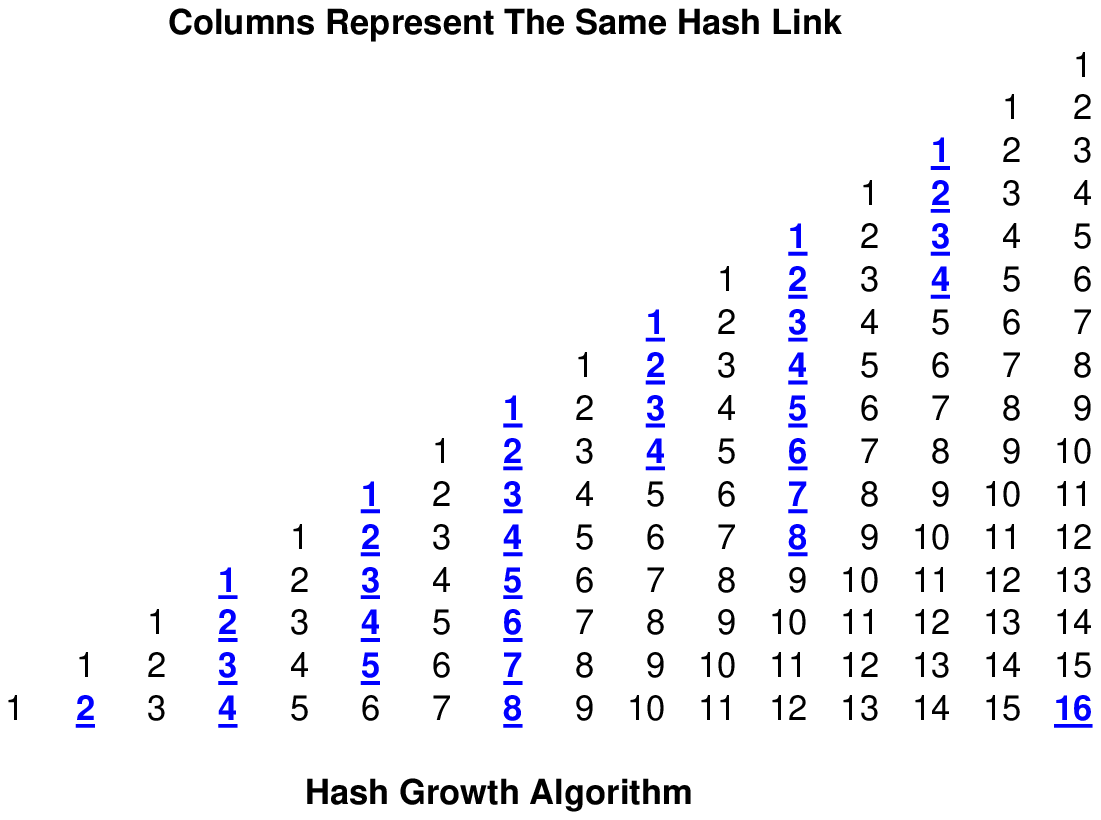}}
\caption{Online Algorithm: As before, each row represents a hash element exposure.  For each row bold and underlined 
positions represent the location of pebbles.}
\label{OLFigure}
\end{figure}

Take $v_i$, then it corresponds to $w_j$, where 

\begin{eqnarray}
j & = & i - (2^{\sigma} - {\bf totalHashElements}),
\label{EQ}
\end{eqnarray}
where $\sigma$ is the total number of pebbles and {\bf totalHashElements} is the number of
hash elements constructed thus far, also ${\bf totalHashElements} \leq 2^{\sigma}$.
Note, that when ${\bf totalHashElements} = 2^{\sigma}$, then we verify $i=j$ by Observation~\ref{Jobserv}
in the appendix.

\begin{fact}
{\sf
Jakobsson's notation and our notation corresponds exactly when $n=2^k$ for 
some integer $k \geq 0$.
}
\label{JFact}
\end{fact}

The online algorithm has two distinct, non-interweavable phases: The growth phase, and the exposure phase.

\subsection{The Growth Phase} 

Next is the growth phase.

\subsubsection{The Online Algorithm's Setup}

	\begin{eqnarray*}
	{\bf seed} 	& \leftarrow & \mbox{the initial hash value}\\
	{\bf totalHashElements} & \leftarrow & 1\\
	{\bf grow.value} & \leftarrow & {\bf seed}\\
	{\bf grow.pebble} & \leftarrow & 1\\
	{\bf exponent }   & \leftarrow & 1
	\end{eqnarray*}

\subsubsection{The Online Algorithm's Init Pebble}

\begin{figure}[h]
\begin{center}
\begin{tabbing}
over over \= text \= text \= text \= text \= text \kill
\> {\bf InitializePebble($p,j, \mbox{\bf grow.value}$)}\\
\> \> $p[j].{\bf Move\_Increment} \leftarrow 2^{j+1}$\\
\> \> $p[j].{\bf Start\_Increment} \leftarrow 3 \times 2^j$\\
\> \> $p[j].{\bf Dest\_Increment} \leftarrow 2^{j+1}$\\
\> \> $p[j].{\bf Position} \leftarrow 2^j$\\
\> \> $p[j].{\bf Destination} \leftarrow 2^j$\\
\> \> $p[j].{\bf Value} \leftarrow {\bf grow.value}$
\end{tabbing}
\end{center}
\caption{Initialize Pebble}
\label{fig:Init}
\end{figure}

\subsubsection{The Online Algorithm's Main} Figure ~\ref{fig:BRalg} updates the pebbles and values.

\begin{figure}[ht]
\begin{center}
\begin{tabbing}
over \=  over \= text \= text \= text \= text \= text \kill
\> \> {\bf while} not done growing hash chain {\bf do}\\
\> \> \> ${\bf grow.value} \leftarrow h({\bf grow.value})$\\
\> \> \> ${\bf totalHashElements} \leftarrow {\bf totalHashElements} + 1$\\
\>1.\> \> {\bf if} ${\bf totalHashElements}-1 = 2^k \mbox{ for some integer $k \geq 1$}$, {\bf then}\\
\> \> \> \> $\mbox{\bf exponent} \leftarrow \mbox{\bf exponent} +1$\\
\> \> \> \> {\bf for} all pebbles $l$ {\bf do}\\
\> \> \> \> \> $p[l].{\bf Position} \leftarrow p[l].{\bf Position} + 2^{\mbox{\bf exponent}-1}$\\  
\> \> \> \> {\bf endfor}\\
\> \> \> \> $j \leftarrow {\bf grow.pebble}$\\
\> \> \> \> {\bf Create} pebble $p[j]$\\
\> \> \> \> {\bf InitializePebble($p,j,\mbox{\bf grow.value}$)}\\
\> \> \> \> $p[j].{\bf distancefromSeed} \leftarrow {\bf totalHashElements}$\\
\> \> \> \> ${\bf grow.pebble} \leftarrow {\bf grow.pebble} + 1$\\
\>2.\> \> {\bf else}\\ 
\> \> \> \> {\bf for} all pebbles $l$ {\bf do}\\
\> \> \> \> \> $t \leftarrow p[l].{\bf Move\_Increment} + p[l].{\bf distancefromSeed} + 1$\\
\>3.\> \> \> \> {\bf if} ${\bf totalHashElements} = t$ {\bf then}\\ 
\> \>3.1\> \> \> \> $p[l].{\bf Value} \leftarrow {\bf grow.value}$\\
\> \>3.2\> \> \> \> $p[l].{\bf distancefromSeed} \leftarrow {\bf totalHashElements}$\\
\> \> \> \> \> \> $p[l].{\bf Position} \leftarrow p[l].{\bf Position} -  p[l].{\bf Move\_Increment}$\\
\> \> \> \> \> {\bf endif}\\
\> \> \> \> {\bf endfor}\\
\> \> \> {\bf endif}\\
\> \> {\bf endwhile}\\
\> \> $j \leftarrow {\bf grow.pebble}$\\
\>4.\> {\bf Create} pebble $p[j]$ where $j \leftarrow {\bf grow.pebble}$\\
\> \> {\bf InitializePebble($p,j,\mbox{\bf seed}$)}\\
\> \> $p[j].{\bf distancefromSeed} \leftarrow 0$\\
\> \> {\bf Sort pebbles by {\bf Position}}
\end{tabbing}
\end{center}
\caption{Hash Chain Growth Algorithm}
\label{fig:BRalg}
\end{figure}

\subsection{The Online Algorithm's Exposure Phase}

In this phase, only the setup is different from Jakobsson's original algorithm, because
we are no longer assuming that exactly $2^j$ pebbles must be present.

\subsubsection{The Online Algorithm's Setup}
	
	{\bf For} $j \leftarrow 1$ to  $\lceil \log n \rceil$ {\bf do}

	\begin{eqnarray*}
	 p[j].{\bf destination} & \leftarrow & p[j].{\bf position}\\
	\end{eqnarray*}

	Furthermore, we must also initialize the next lines
	
	\begin{eqnarray*}
	{\bf current.position} & \leftarrow & {\bf totalHashElements}\\
	{\bf current.value} & \leftarrow & {\bf grow.value}
	\end{eqnarray*}

%
%
%
%

The online algorithm's Main function is performed exactly as Jakobsson's algorithm with the exception
that {\bf current.position} cannot simply be set to zero during the setup phase, but rather must be determined from 
the value of the first pebble in the sorted order.

\subsection{Characteristics of the Hash Chain Growth Algorithm}

To prove the validity of the Hash Chain Growth Algorithm we will show: (1) at each 
step of the hash chain growth there are always enough pebbles, and (2) the pebbles
are always properly placed such that Jackobsson's algorithm can immediately begin
to run on the stored data from the generated hash chain.

Recall, given $n=2^t$, for some integer $t$, then pebbles are placed in hash element positions 
\begin{eqnarray*}
t = \log n, \ \ t-1 = \log (n/2), \ \ t-2 = \log (n/4), \ \ \cdots, \ \ t=1.
\end{eqnarray*}

For any initial pebble $p[j]$ the value $p[j].{\bf Destination}$ never
decreases in Jakobsson's algorithm.
Thus, to determine on which move back the initial pebble $p[j]$ will be output, compute when
\begin{eqnarray*}
p[j].{\bf Destination} > n
\end{eqnarray*}
first occurs. 

Assume $n$ is a power of $2$, and the definition of $D(j,k)$ can be found in the appendix, and given a fixed $j$
solve for the minimal $k$ so that
\begin{eqnarray*}
2^j + k \cdot 2^{j+1} & > & 2^{\log n}\\
2k & > & 2^{\log n - j} -1.
\end{eqnarray*}
For instance, for $j = \log (n/2)$, then $k = 1$ gives
\begin{eqnarray*}
2k  = 2 & > & 2^{\log n - (\log n -1)} -1\\
	& > & 2 -1 \ \ = \ \ 1,
\end{eqnarray*}
which holds.

\begin{fact}
{\sf
Let $n = 2^k$.
In Jakobsson's algorithm, pebbles are eliminated in the reverse order of their original indices 
$j = k-1, \cdots, 1$ where we consider the seed pebble at $j = k$ to never be discarded. 
}
\label{fact12}
\end{fact}

\begin{proof}
This proof follows from Jakobsson's algorithm in Figure~\ref{Fig:JsAlgorithm}.  
By Fact~\ref{F5} of the appendix, it must be that only the pebble in initial position 
$j=k-1=\log(n/2)$ is going to be eliminated
immediately after Jakobsson's algorithm generates and emits $n/2$ hash elements. 

Now, suppose the hash elements are renumbered where $1$ represents hash element $n/2+1$; $2$ represents
hash element $n/2+2$, all the way to $n/2$ represents hash element $n$. Still $n = 2^k$.

By Fact~\ref{F9} of the appendix, with this re-numbering, the remaining pebbles will be placed at pebble positions
\begin{eqnarray*}
2^1, 2^2, \ \cdots, \ 2^{k-1},
\end{eqnarray*}
where $n/2 = 2^{k-1}$. The pebble in position $p[k-1]$ is the seed of the hash chain, so it never moves.
By Corollary~\ref{C1} also of the appendix, for each of these remaining pebbles,
\begin{eqnarray*}
p[j].{\bf Position} & = & p[j].{\bf Destination}.
\end{eqnarray*}

However, for any initial pebble $p[j]$ the value $p[j].{\bf Destination}$ never
decreases in the algorithm of Figure~\ref{Fig:JsAlgorithm}.
Thus, we need to determine what initial pebble $p[j]$ is placed in position $2^{k-2} = n/4$.
This new position is one backward move beyond the $n/2$nd emitted hash element.
Thus, the ${\bf Dest\_Increment}$ of this pebble must be $2^{k-2} = n/4$, so
the initial pebble was $p[k-2]$. Furthermore, by Fact~\ref{F5}, this pebble is the second
pebble to be discarded.

The proof is completed by induction.
\end{proof}

The next facts follow from the online algorithm in Figure~\ref{fig:BRalg}.

\begin{fact}
{\sf
Consider the online hash growth algorithm (see Figure~\ref{fig:BRalg}).
A new pebble is only created when {\bf totalHashElements} is such that
\begin{eqnarray*}
	2^{k}+1 = {\bf totalHashElements},
\end{eqnarray*}
\noindent
for some $k \geq 1$.
}
\label{olf1}
\end{fact}

\begin{proof}
This is a direct result of line~1 of the online hash growth algorithm.
\end{proof}

Any time $2^{k} = {\bf totalHashElements} -1$, then between now and when
$2^{k-1} = {\bf totalHashElements} -1$, all of the pebble's positions must be shifted
by $2^{k-1}$. This is done in Figure~\ref{fig:BRalg} before a new pebble is created.

\begin{fact}
{\sf
Consider the online hash growth algorithm (see Figure~\ref{fig:BRalg}).
Each time an element is added to the chain, {\bf totalHashElements} is incremented by one.
}
\label{olf2}
\end{fact}

\begin{fact}
{\sf
Consider the online hash growth algorithm (see Figure~\ref{fig:BRalg}).
Each time a pebble is created it is initialized as,
\begin{eqnarray*}
p[j].{\bf distancefromSeed} & \leftarrow & {\bf totalHashElements}.
\end{eqnarray*}
}
\label{olf3}
\end{fact}

This follows directly from section~1 of Figure~\ref{fig:BRalg}.

\begin{fact}
{\sf
Consider the online hash growth algorithm (see Figure~\ref{fig:BRalg}).
After its initialization, the only time $p[j].{\bf distancefromSeed}$ changes is when 
pebble $j$ is moved and $p[j].{\bf distancefromSeed}$ is always
updated to maintain the correct distance from pebble $p[j]$ to the hash chain seed.
}
\label{olf8}
\end{fact}

\begin{proof}
After $p[j].{\bf distancefromSeed}$'s initialization in section~1 of Figure~\ref{fig:BRalg},
the value $p[j].{\bf distancefromSeed}$ only changes
when the next assignment is made in line~3.2:

\begin{eqnarray*}
 p[j].{\bf distancefromSeed} & \leftarrow & {\bf totalHashElements}
\end{eqnarray*}

and this occurs exactly when the condition in line~3 holds true since
\begin{eqnarray*}
 p[j].{\bf distancefromSeed} +  p[j].{\bf Move\_Increment} & = & {\bf totalHashElements},
\end{eqnarray*}
completing the proof.
\end{proof}

\begin{fact}
{\sf
In order to reverse Jakobsson's algorithm pebbles should be assigned indices in ascending order. 
}
\label{olf4}
\end{fact}

\begin{proof}
This holds by Fact~\ref{fact12}.
\end{proof}

\begin{fact}
{\sf
For each pebble, to find the current position for the purposes of the following proof we can calculate,
\begin{eqnarray*}
	p[j].{\bf Position} = {\bf totalHashElements} - p[j].{\bf distancefromSeed}.
\end{eqnarray*}
}
\label{olf5}
\end{fact}

\begin{proof}
This holds because {\bf Position} is a measure of how far from the front of the chain a pebble currently is.  By Fact~\ref{olf2} 
{\bf totalHashElements} is always equal to the total number of elements in the chain and by Facts~\ref{olf3} and~\ref{olf5} {\bf distancefromSeed} 
is always the distance the pebble is from the seed including the seed itself.  Thus, we know that we can find how far from the 
front of the hash chain ($p[j].{\bf Position}$) a pebble is by ${\bf totalHashElements} - p[j].{\bf distancefromSeed}$.
\end{proof}

\begin{fact}
{\sf
Consider the online hash growth algorithm.
The only time a pebble is moved is when ${\bf totalHashElements} = t$ 
where $t \leftarrow p[j].{\bf Move\_Increment} + p[j].{\bf distancefromSeed}$. 
When a pebble is moved, then it is moved to the front of the chain.
}
\label{olf7}
\end{fact}

\begin{fact}
{\sf
When the online hash growth algorithm is halted the final pebble is placed at the root with the 
seed value, and the next available pebble index.
}
\label{olf6}
\end{fact}

Fact~\ref{olf6} follows immediately from the code starting at line~4 of Figure~\ref{fig:BRalg}.
Jakobsson assumes $n$ is a power of two for simplicity~\cite{J}. In the case of the online algorithm, 
values of $n$ other than powers of two are also critical.

\begin{theorem}
{\sf
Suppose that the online algorithm in Figure~\ref{fig:BRalg} is halted
after generating $n$ total hash elements.  The pebbles associated with the hash chain of size $n$ 
will be stored at positions equivalent to the positions in Jakobsson's algorithm.
}
\end{theorem}

\begin{proof}
The case where $n=2^k$, the online and amortized algorithm both have pebbles in the same positions
by Fact~\ref{JFact}.

The proof follows by induction on the ranges from $[2^k, 2^{k+1})$ to $[2^{k+1}, 2^{k+2})$, for $k \geq 2$.\\

\noindent
{\bf Basis:} Take a hash chain with $k = 2$, thus $n = 2^2 = 4$ elements as stored by the online
by the algorithm in Figure~\ref{fig:BRalg}. 
\begin{eqnarray*}
p[1].{\bf Position} & = & 2\\
p[1].{\bf Move\_Increment} & = & 4.
\end{eqnarray*}

As the hash chain is being grown after $n=4$, the first occurrence of {\bf CreatePebble} is at $2^2 + 1 = 5$.  
This pebble is placed at the front of the chain by Fact~\ref{olf3} and labeled $p[2]$ by Fact~\ref{olf4}.
Furthermore,
\begin{eqnarray*}
p[2].{\bf Position} & = & 4\\
p[2].{\bf Move\_Increment} & = & 8.
\end{eqnarray*}
since ${\bf grow.pebble} = 2$.
Suppose the hash chain growth is halted at $n=5$.  Due to Fact~\ref{olf6} we know that the final pebble
is placed for the seed. The pebbles are sorted by their {\bf Position}s.
  
Growing the hash chain to $n=6$, prompts the online hash growth algorithm to check each pebble
for ${\bf totalHashElements} = t = 6$ where,
\begin{eqnarray*}
t & \leftarrow & p[j].{\bf Move\_Increment} + p[j].{\bf distancefromSeed}.
\end{eqnarray*}
But, ${\bf totalHashElements} \neq t$ for all $j$ and $n = 6$.  

The algorithm finds $t_j$ where 
\begin{eqnarray*}
t_1 & \leftarrow & p[1].{\bf Move\_Increment} + p[1].{\bf distancefromSeed}\\
t_2 & \leftarrow & p[2].{\bf Move\_Increment} + p[2].{\bf distancefromSeed}.
\end{eqnarray*}

At $n=5$ no pebble movement occurs.  At $n = 6 = t_1$ the algorithm sets 
\begin{eqnarray*}
p[1].{\bf distancefromSeed} & = & p[1].{\bf distancefromSeed} + p[1].{\bf Move\_Increment},
\end{eqnarray*}
such that 
\begin{eqnarray*}
p[1].{\bf distancefromSeed} & = & {\bf totalHashElements} \ \ = \ \ 6.  
\end{eqnarray*}
Suppose the hash chain growth is halted at $n=7$. Again, due to Fact~\ref{olf6} we know that the final pebble
is the seed.
Thus, for $n = 7$ we have pebbles at $p[1].{\bf Position} = 2$, $p[2].{\bf Position} = 4$
and $p[3].{\bf Position} = 8$ by 
Fact~\ref{olf5}.\\

\noindent
{\bf Inductive Hypothesis:} The statement of this Lemma holds for 
hash chains from $n=2^{k}$ to $n=2^{k+1}$, for some $k \geq 2$. \\

\noindent
{\bf Inductive Step:} 
Consider the case where $n$ is increased from $n=2^{k}$ to $n=2^{k+1}$ where $k \geq 2$. By the inductive hypothesis
we suppose the pebbles are all placed in the correct positions for $n \leq 2^{k}$.

By Jakobsson's algorithm, for a chain of original size $n$ we know that a pebble is only discarded when the 
total number of remaining elements is $n/2$.  Thus, the hash chain growth algorithm only adds new elements when the 
total number of pebbles is $(n/2)+1$.  

Recall, a pebble is always placed at the seed starring at line~4 of the online algorithm, this assures we 
always have the proper number of pebbles.

By Fact~\ref{F1} we know that in Jakobsson's algorithm only pebble $p[1]$ is moved.  Further, it is always 
moved a total of $p[1].{\bf Dest\_Increment}$ 
after being moved such that it reaches its destination.  Since 
\begin{eqnarray*}
p[i].{\bf Dest\_Increment} & = & p[i].{\bf Move\_Increment}
\end{eqnarray*}
and because pebbles are always moved to the front by the online algorithm by Fact~\ref{olf7},
then we know that pebbles are always moved correctly.  

For each $n$, there are always 
enough pebbles to support the total number of elements emitted in the online algorithm. 
As just shown, these pebbles are always placed correctly completing the proof. 
\end{proof}

\section{Further Directions}
\label{Future}

It would be interesting to extend the algorithms of Sella~\cite{Sella} or Matias and Porat~\cite{MP} 
to work online. That is, suppose we initially have a data structure capable of holding $n$ points of data,
then it would be interesting to be able to generate $n+1$ elements of data and expand
this data structure suitably, i.e. not doubling it in size.

We have not addressed the tamper-resistant hardware. Likewise, for very small time granularity, we have
not addressed clock drift and related timing issues.

\section{Acknowledgements}

Thanks to Markus Jakobsson for his comments on this work.
Additionally, the referees' comments were very helpful and insightful.
Thanks to Judge Joseph Colquitt for his perspective on chains of custody and providing us with~\cite{Colquitt}.

\bibliographystyle{plain}%

\pagebreak

\section*{Appendix: A Proof of Correctness for Jakobsson's Algorithm}

This appendix gives a correctness proof for Jakobsson's amortized algorithm. This is done
to derive the correctness of the new online algorithm given in this paper.  
Jakobsson's paper~\cite{J} does not give a complete proof of correctness for his algorithm. 
However, our online result seems to require certain factual statements from the correctness
of Jakobsson's algorithm.

Ultimately, what we show when Jakobsson's algorithm runs emitting $n/2 = 2^{k}$ hash elements
the remaining pebbles contain hash elements of hash-chain positions 
$2^1, 2^2, 2^3, \ldots, 2^k$ which will prove the correctness of Jakobsson's algorithm.

Jakobsson's work immediately gives the next observation.

\begin{observation}
{\sf
Suppose a hash chain has $n$ hash elements and $n = 2^k$, for some integer $k \geq 1$
and Jakobsson's algorithm is about to start.
The $k$ pebbles initially contain hash elements in hash-chain positions 
\begin{eqnarray*}
2^1, 2^2, 2^3, \cdots, 2^k.
\end{eqnarray*}
Say Jakobsson's algorithm is about to start on a hash chain of size $2n$.
Then $k+1$ pebbles initially contain hash elements in hash-chain positions 
\begin{eqnarray*}
2^1, 2^2, 2^3, \cdots, 2^k, 2^{k+1}.
\end{eqnarray*}
}
\label{Jobserv}
\end{observation}

A proof of Observation~\ref{Jobserv} follows directly from Jakobsson's setup phase where the pebbles are initialized.

\begin{definition}
{\sf
A pebble $p[j]$ is moved {\em backward} if $p[j].{\bf Position}$ is increased in line~3.1 of the algorithm in 
Figure~\ref{Fig:JsAlgorithm}.
A pebble $p[j]$ is moved {\em forward} if $p[j].{\bf Position}$ is decreased by~2 in line~2.1 of this algorithm.
}
\end{definition}

\begin{fact}
{\sf
Only pebble $p[1]$ moves backwards in Jakobsson's algorithm. Furthermore, it only does so when {\bf Current.position} is even.
}
\label{F1}
\end{fact}

\begin{proof}
This holds since, the major else in section~3 of Figure~\ref{Fig:JsAlgorithm} is the only place Jakobsson's algorithm
increases any pebble position.
\end{proof}

\begin{fact}
{\sf
A pebble $p[j]$ moves two hash elements forward in lines {\rm 2.1} and {\rm 2.2} of Jakobsson's algorithm only after $p[j]$ was moved previously from the front of
the chain by Fact~\ref{F1}.
}
\label{F2}
\end{fact}

\begin{proof}
In code section~2, making $p[j].{\bf Position} \neq p[j].{\bf Destination}$ may only be done by code in section~3
of Figure~\ref{Fig:JsAlgorithm}.
\end{proof}

Recall that Jakobsson's algorithm initially sets $p[j].{\bf Position} = p[j].{\bf Destination}$, for all pebbles $j: \sigma \geq j \geq 1$.
Where $\sigma$ is the number of pebbles.

\begin{fact}
{\sf
For any pebble $p[j]$, before or after any invocation of Jakobsson's algorithm in Figure~\ref{Fig:JsAlgorithm}, the bound holds:
\begin{eqnarray*}
p[j].{\bf Position} & \geq & p[j].{\bf Destination},
\end{eqnarray*}
for all $j: \sigma \geq j \geq 1$ where $p[j].{\bf Destination} \neq +\infty$.
}
\label{F3}
\end{fact}
\begin{proof}
For each pebble $j \geq 1$, the values ${\bf Start\_Increment} = 3 \times 2^{j}$ and 
${\bf Dest\_Increment} = 2^{j+1}$ are both even and never change.
Initially in the setup phase
\begin{eqnarray*}
p[j].{\bf Destination} & = & 2^j\\
p[j].{\bf Position} & = & 2^j.
\end{eqnarray*}

In section~3, the bound $p[j].{\bf Position} \geq p[j].{\bf Destination}$ is always
enforced when {\bf Current.position} is even, due to the values of ${\bf Start\_Increment}$
and ${\bf Dest\_Increment}$ and also the assignments on lines~3.1 and~3.2,
\begin{eqnarray*}
p[j].{\bf Position} & \leftarrow & p[j].{\bf Position} + p[j].{\bf Start\_Increment}\\
p[j].{\bf Destination} & \leftarrow & p[j].{\bf Destination} + p[j].{\bf Dest\_Increment}.
\end{eqnarray*}

However, in section~2 of Figure~\ref{Fig:JsAlgorithm},
if $p[j].{\bf Destination} \neq p[j].{\bf Position}$, then $p[j].{\bf Position}$
is decremented as follows:
\begin{eqnarray*}
p[j].{\bf Position} & \leftarrow & p[j].{\bf Position}  -2.
\end{eqnarray*}

By Fact~\ref{F2}, during a run of Jakobsson's algorithm, both
\begin{eqnarray*}
p[1].{\bf Position} \mbox{ and } p[1].{\bf Destination}
\end{eqnarray*}
remain even values since lines~3.1 and~3.2 add even values to even values.

Line~3.2 ensures 
\begin{eqnarray*}
p[1].{\bf Position} & \geq & p[1].{\bf Destination}
\end{eqnarray*}
and line~2.1 decrements $p[j].{\bf Position}$ by~2, spanning all even numbers going down.
Thus $p[j].{\bf Position}$ must
eventually equal $p[j].{\bf Destination}$, completing the proof.
\end{proof}

\begin{definition}
{\sf
The expression $p[{\bf j} \overset{k}{\rightarrow} i]$ indicates the pebble initially setup in position $j$ is
currently in position $i$ after $k$ backward moves.
The expression $p[j \overset{k}{\rightarrow} i]$ indicates this pebble was previously in position $j$, but $j$ is
not necessarily the initial position of this pebble in setup.
}
\end{definition}

The expression $p[j \overset{0}{\rightarrow} 1]$ indicates a pebble that was in position $j$ becomes $p[1]$ without
moving backwards. That is, the pebble labeled $p[j]$ 
becomes $p[1]$ after emission of $j-1$ hash elements. 

Note, $p[{\bf j} \overset{k}{\rightarrow} 1]$ indicates that the pebble initially
labeled $p[j]$ has become pebble $p[1]$ exactly $k$ times.
Also, $p[{\bf 1} \overset{1}{\rightarrow} j]$ indicates that the unique pebble initially 
labeled $p[1]$ is moved backwards exactly once and eventually relabeled pebble $p[j]$.
Where $p[1 \overset{1}{\rightarrow} j]$ is a pebble that was earlier in position~1 and moved back once,
but perhaps it was in another position initially at setup.

\begin{definition}
{\sf
Let $D(j,k)$ be the value of $p[{\bf j} \overset{k}{\rightarrow} i].{\bf Destination}$ immediately after $k$ {\em backward moves} of the pebble initially setup as pebble $p[j]$ and became $p[i]$, by the Algorithm in Figure~\ref{Fig:JsAlgorithm}.
}
\end{definition}

Thus, by Fact~\ref{F1}, each backward move is initiated when the initial $p[j]$ has become $p[1]$.

Next is a bound on $D(j,k)$, for the initial pebble $p[j]$:
\begin{eqnarray*}
D(j,k) & = &
	\left\{
	\begin{array}{ll}
	2^j+2^{j+1} & \mbox{ if } k=1\\
	D(j,k-1)+2^{j+1} & \mbox{ otherwise. }
	\end{array}
	\right.
\end{eqnarray*}

\begin{definition}
{\sf
Let $P(j,k)$ be an upper-bound on the value of $p[{\bf j} \overset{k}{\rightarrow} i].{\bf Position}$ immediately after $k$ {\em backward moves} of the pebble initially setup as pebble $p[j]$ and became $p[i]$, by the Algorithm in Figure~\ref{Fig:JsAlgorithm}.
}
\end{definition}

Next is a bound on $P(j,k)$, for the initial pebble $p[j]$:

\begin{eqnarray*}
P(j,k) & \leq &
	\left\{
	\begin{array}{ll}
	2^j+ 3 \cdot 2^{j} & \mbox{ if } k=1\\
	P(j,k-1)+ 3 \cdot 2^{j} & \mbox{ otherwise. }
	\end{array}
	\right.
\end{eqnarray*}

\begin{fact}
{\sf
Immediately after moving pebble $p[1]$ backwards to $p[i]$, the next upper bound holds:
\begin{eqnarray*}
p[{\bf j} \overset{k}{\rightarrow} i].{\bf Position} - p[{\bf j} \overset{k}{\rightarrow} i].{\bf Destination} 
& \leq & 2^j + k \left( 3 \cdot 2^j \right) - 2^j - k \left( 2 \cdot 2^j \right)\\
& \leq & 
k \cdot 2^j.
\end{eqnarray*}
Equality holds immediately after $p[{\bf j} \overset{k}{\rightarrow} i]$ is placed in position $i$.
}
\label{F3-5}
\end{fact}

This fact holds by Fact~\ref{F3}.

\begin{fact}
{\sf
Consider the Algorithm in Figure~\ref{Fig:JsAlgorithm}.
Then for every pebble $p[{\bf j} \overset{k}{\rightarrow} 1]$, it must be that
\begin{eqnarray*}
p[{\bf j} \overset{k}{\rightarrow} 1].{\bf Position} & = & p[{\bf j} \overset{k}{\rightarrow} 1].{\bf Destination},
\end{eqnarray*}
where neither
$p[{\bf j} \overset{k}{\rightarrow} 1].{\bf Position} = +\infty$ nor
$p[{\bf j} \overset{k}{\rightarrow} 1].{\bf Destination} = +\infty$.
}
\label{F4}
\end{fact}

\begin{proof}
Consider the next function that describes how $p[j].{\bf Destination}$ works.
Note, that line~3.2 of Figure~\ref{Fig:JsAlgorithm} is the only place the {\bf Destination} field is changed.\\

\noindent
{\bf Basis:} It must be that
\begin{eqnarray*}
p[{\bf j} \overset{1}{\rightarrow} 1].{\bf Position} & = & p[{\bf j} \overset{1}{\rightarrow} 1].{\bf Destination},
\end{eqnarray*}
since by the upper-bound in Fact~\ref{F3-5},
\begin{eqnarray*}
p[{\bf j} \overset{1}{\rightarrow} 1].{\bf Position} - p[{\bf j} \overset{1}{\rightarrow} 1].{\bf Destination} 
& \leq &
2^j.
\end{eqnarray*}
However, $D(j,1) = 2^j+2^{j+1}$ and by line~2.1 of the algorithm, we must decrement
$p[j \overset{1}{\rightarrow} 1].{\bf Position}$ by 2 each of $D(j,1)$ times. This is because
the pebbles are sorted by {\bf Position}. So immediately after the initial pebble $p[j]$ is moved
back, then its destination is $D(j,1)$. Thus, after $D(j,1)$ more hash elements are emitted,
then this pebble will be labeled $p[1]$.

Thus, since $2D(j,1) > 2^j$, by Fact~\ref{F3}, 
\begin{eqnarray*}
p[{\bf j} \overset{1}{\rightarrow} 1].{\bf Position} & = & p[{\bf j} \overset{1}{\rightarrow} 1].{\bf Destination},
\end{eqnarray*}
completing the basis.\\

\noindent
{\bf Inductive Hypothesis:} Suppose in $k \geq 1$ backward moves of the initial pebble $p[j]$ by the Algorithm in Figure~\ref{Fig:JsAlgorithm}, we have
\begin{eqnarray*}
p[{\bf j} \overset{k}{\rightarrow} 1].{\bf Position} & = & p[{\bf j} \overset{k}{\rightarrow} 1].{\bf Destination}.
\end{eqnarray*}

\noindent
{\bf Inductive Step:} For some $k \geq 1$, consider $k + 1$ backward moves of the initial pebble $p[j']$ by Algorithm in Figure~\ref{Fig:JsAlgorithm}, and this step will show:
\begin{eqnarray*}
p[{\bf j'} \overset{k+1}{\longrightarrow} 1].{\bf Position} & = & p[{\bf j'} \overset{k+1}{\longrightarrow} 1].{\bf Destination}.
\end{eqnarray*}

Start by considering $k+1$ backward moves of the initial pebble $p[j']$, for the sake of a
contradiction suppose,
\begin{eqnarray*}
p[{\bf j'} \overset{1}{\rightarrow} j \overset{k}{\rightarrow} 1].{\bf Position} 
& \neq & 
p[{\bf j'} \overset{1}{\rightarrow} j \overset{k}{\rightarrow} 1].{\bf Destination},
\end{eqnarray*}
and in any case, by Fact~\ref{F3-5}, the next bound holds:
\begin{eqnarray*}
p[{\bf j'} \overset{1}{\rightarrow} j].{\bf Position} - p[{\bf j'} \overset{1}{\rightarrow} j].{\bf Destination}
& \leq &
2^{j'}.
\end{eqnarray*}

However $D(j',1) = 2^{j'+1} + 2^{j'}$, which means
after $2^{j'+1} + 2^{j'}$ hash elements are emitted to take the initial pebble
$p[j']$ to $p[1]$ the first time, then 
\begin{eqnarray*}
p[{\bf j'} \overset{1}{\rightarrow} j].{\bf Position}
\end{eqnarray*}
must be decremented by~2 in each run of section~2 of
Figure~\ref{Fig:JsAlgorithm} decreasing $p[{\bf j'} \overset{1}{\rightarrow} 1].{\bf Position}$ 
by at least $2 D(j',1) > 2^{j'}$. 

Now, by the Inductive Hypothesis it must be that,
\begin{eqnarray*}
p[j \overset{k}{\rightarrow} 1].{\bf Position} & = & p[j \overset{k}{\rightarrow} 1].{\bf Destination}.
\end{eqnarray*}
giving,
\begin{eqnarray*}
p[{\bf j'} \overset{k+1}{\longrightarrow} 1].{\bf Position}
& = & p[{\bf j'} \overset{k+1}{\longrightarrow} 1].{\bf Destination},
\end{eqnarray*}
indicating our supposition was incorrect, 
which completes the proof.
\end{proof}

\begin{fact}
{\sf
Suppose a hash chain starts with $n = 2^{k+1}$ total hash elements, for some integer $k \geq 1$
and Jakobsson's algorithm runs emitting $n/2 = 2^{k}$ hash elements.
Then the pebble initially in position $j < k$ is moved backwards a total of 
\begin{eqnarray*}
2^{k-j-1}
\end{eqnarray*}
times.
}
\label{F7}
\end{fact}

\begin{proof}
By Fact~\ref{F1}, a pebble only moves back when it is in pebble position~1.
By Fact~\ref{F3},
\begin{eqnarray*}
p[j].{\bf Position} & \geq & p[j].{\bf Destination},
\end{eqnarray*}
so the only concern is how $p[j].{\bf Destination}$ is initialized and how $p[j].{\bf Destination}$ changes.

After $n/2$ hash elements are emitted, it takes exactly 
\begin{eqnarray*}
\left( \frac{n}{2} - 2^{j} \right) \frac{1}{p[j].{\bf Dest\_Increment} }
& = &
\left( 2^{k} - 2^{j} \right) \frac{1}{2^{j+1}}\\
& = & 
2^{j-k-1} - \frac{1}{2},
\end{eqnarray*}
increments of $p[j].{\bf Destination}$ by the time the pebble initially at position $j$ is disposed of.
\end{proof}

A pebble is disposed of when $p[j].{\bf Destination} \leftarrow +\infty$.
In terms of Fact~\ref{F7}, the pebble in position $k+1$ is never moved back and the pebble in position $k$ is disposed
of when it is moved back the first time.

\begin{fact}
{\sf
Given an $n =2^k$ element hash chain and take the Algorithm in Figure~\ref{Fig:JsAlgorithm}.
The initial pebble in position $j$, for $j=k-1 = \log (n/2)$ is the only pebble that is disposed
of by the time the first $n/2$
hash elements are emitted.
}
\label{F5}
\end{fact}

\begin{proof}
For each pebble, the values {\bf Start\_Increment} and {\bf Dest\_Increment} never change.
Initially in setup $p[j].{\bf Destination} \leftarrow 2^j$, for all $j: \log n \geq j \geq 1$.

The value $p[j].{\bf Destination}$ is only increased in line~3.2 of Figure~\ref{Fig:JsAlgorithm}.
Considering line~3.2 gives the bound
\begin{eqnarray*}
p[j].{\bf Destination} & = & 2^j + p[j].{\bf Dest\_Increment}\\
p[j].{\bf Destination} & = & 2^j + 2^{j+1}\\
p[j].{\bf Destination} & = & 3 \cdot 2^{j}.
\end{eqnarray*}

Pebble $p[1]$ is the only one that is moved back by Fact~\ref{F1}. Further,
\begin{eqnarray*}
3 \cdot 2^{j} & > & n,
\end{eqnarray*}
holds for $j \geq \log(n/2)$. But the only mobile pebble that satisfies this constraint is
initially in position $j=\log(n/2) = k-1$.
There is a pebble $j=\log n$ at the seed of the hash chain, but this never moves. 

Now, take any initial pebble that started in position $j \leq \log(n/4)$ then by Fact~\ref{F7},
such a pebble makes $2^{k-j-1}$ backward moves by the time $n/2$ hash elements are emitted.
Thus, by the time the $n/2$-nd pebble is emitted, the initial pebble $p[j]$ is moved
from the starting position $2^j$ to position
\begin{eqnarray*}
2^{j} + \left( 2^{j+1} \cdot 2^{k-j-1} \right)  & = & 2^j + 2^k,
\end{eqnarray*}
and the largest $j$ and $k$ may be is $\log (n/4)$ and $\log (n/2)$, respectively, giving

\begin{eqnarray*}
2^j + 2^k  	& \leq & 2^{\log n/4} + 2^{\log n/2}\\
		& \leq & \frac{3n}{4} \ \ < n,
\end{eqnarray*}
completing the proof.
\end{proof}

\begin{fact}
{\sf
Suppose a hash chain starts with $n = 2^{k+1}$ total hash elements, and for some integer $k \geq 1$,
while Jakobsson's algorithm runs emitting the first $n/2 = 2^{k}$ hash elements,
let $i$ be the first placement of $p[{\bf j} \rightarrow i]$ between hash elements
$2^k$ and $2^{k+1}$.
Then, Jakobsson's algorithm will
emit a total of 
\begin{eqnarray*}
\frac{n}{2} - \left( 2^{k-j-1} -1 \right) 2^{j+1} - 2^{j}
\end{eqnarray*}
hash elements before emitting hash element $2^k$ but immediately
after $p[{\bf j} \rightarrow i]$ is placed at position~$i$.
}
\label{F8}
\end{fact}

\begin{proof}
After Jakobsson's algorithm emits hash element $2^k$, a total of $n/2$ total hash 
elements have already been emitted.

By Fact~\ref{F7} the pebble initially at position $j < k$ is moved backwards
$2^{k-j-1}$ times. That is, it appears in pebble position~1, a total of $2^{k-j-1}$ times.

The last time this pebble moves backward to position $i$ is the first time it moves to 
a position behind hash position $2^k$. This gives the factor of $(2^{k-j-1} -1)$.

Finally, for initial pebble $j$, since {\bf Dest\_Increment} is always $2^{j+1}$,  
each time we move backwards in Jakobsson's algorithm, we move {\bf Destination}
back by $2^{j+1}$ elements. 
That is, we move back a total of 
\begin{eqnarray*}
\left( 2^{k-j-1} -1 \right) 2^{j+1}
\end{eqnarray*}
hash elements by the time the pebble initially labeled $j$ passes hash
position $2^k$.

Furthermore, the initially labeled $j$th pebble
starts at position $2^j$, so we adjust for the number of hash elements emitted it takes to
traverse $p[{\bf j} \overset{0}{\rightarrow} 1]$.
\end{proof}

\begin{corollary}
{\sf
We know that when $n/2$ hash elements are emitted that all remaining pebbles will have
\begin{eqnarray*}
p[j].{\bf Destination} & = & p[j].{\bf Position}. 
\end{eqnarray*}
}
\label{C1}
\end{corollary}
\begin{proof}
We know by Fact~\ref{F8}, that there are
\begin{eqnarray*}
\frac{n}{2} - \left( 2^{k-j-1} -1 \right) 2^{j+1} -2^{j}
\end{eqnarray*}

\noindent
elements that remain to be output before a total of $n/2$ elements are emitted after 
each pebble at original pebble index $j$ is moved for the last time before $n/2$ elements are
emitted.

By Jakobsson's algorithm we know that each time a pebble at original position $j$ is moved, 
it will take,
\begin{eqnarray*}
(3 \cdot 2^j - 2^{j+1})/2
\end{eqnarray*}

\noindent
hash exposures for $p[j].{\bf Position} = p[j].{\bf Destination}$.

Thus, since
\begin{eqnarray*}
\frac{n}{2} - \left( 2^{k-j-1} -1 \right) 2^{j+1} -2^{j} > (3 \cdot 2^j - 2^{j+1})/2
\end{eqnarray*}

\noindent
we know that when $n/2$ hash elements are emitted that all remaining pebbles will have
$p[j].{\bf Position} = p[j].{\bf Destination}$.  Thus, completing the proof.
\end{proof}

The next fact, can also serve as an alternate proof-of-correctness for Jakobsson's
algorithm~\cite{J}. It is presented here, since it is needed subsequently.

First, the function $R()$ reindexes the pebble indices from one power 
of two down to the next lowest power of two. When $n/2$ pebbles are emitted, then the total number
of remaining hash elements goes from $n = 2^{k+1}$ down to $n = 2^{k}$. 
\begin{eqnarray*}
R_{k}(l_{k+1}) = l_{k+1} - 2^k,
\end{eqnarray*}

\noindent
where $l_{k+1}$ is the hash element $\geq 2^{k+1}$ contained by pebble index $k+1$, 
and $R_k(l_{k+1})$ is the adjusted index of the hash element $\geq 2^k$.

\begin{fact}
{\sf
Suppose a hash chain starts with $n = 2^{k+1}$ total hash elements, for some integer $k \geq 1$
and Jakobsson's algorithm runs emitting $n/2 = 2^{k}$ hash elements.
Then the remaining pebbles contain hash elements of hash-chain positions 
\begin{eqnarray*}
2^1, 2^2, 2^3, \cdots, 2^k.
\end{eqnarray*}
}
\label{F9}
\end{fact}

\begin{proof}
This proof follows by induction on the algorithm in Figure~\ref{Fig:JsAlgorithm}.
Setting $k=2$ ensures sections~1,2 and~3 of 
Jakobsson's algorithm will run (see Figure~\ref{Fig:JsAlgorithm}).  The base case will
be $k=2$, so $n=2^3=8$, exposing $n/2$ elements to get $k-1$, $n=2^{k-1}$ total 
elements.
\\

\noindent
{\bf Basis:} Take a hash chain with $n=2^{3}$ total hash elements.
Thus, $n=2^{k+1}$, so $k=2$.
Then the setup phase of Jakobsson's algorithm assigns pebbles to hold
hash elements in positions $2^1, 2^2$ and $2^3$.

Now, suppose $n/2 = 4$ hash elements are exposed, thus the algorithm in Figure~\ref{Fig:JsAlgorithm}
is called 4~times. 

By Fact~\ref{F5} only one pebble, the pebble associated with the $2^k$ position will be discarded.
This leaves $2^{k+1}-1 = 2$ pebbles remaining.  One of these pebbles is 
associated with the root and will not change.  For the remaining pebble associated with the
$2^1$ position

\begin{eqnarray*}
P(j,1) & = & 2^j+ 3 \cdot 2^{j} = 2^1+ 3 \cdot 2^{1} = 8\\
D(j,1) & = & 2^j+ 2^{j+1} = 2^1+ 2^2 = 6.
\end{eqnarray*}

Now, $n=2^k=4$ elements remain.  In order to put position $l$, which is in terms of base $2^{k+1}$ 
hash chain, in terms of the base $2^k$ hash chain we have
our two remaining pebbles at $R_{k=2}(6_{g=3})=2$, and $R_{k=2}(8_{g=3})=4$, where $g=k+1$. Since $k=2$ our remaining pebbles
are at positions $2^1$ and $2^2$ or, $2^1 \ldots 2^k$.\\

\noindent
{\bf Inductive Hypothesis:} Suppose that for a hash chain of size $2^{k+1}$ where
$k \geq 2$ after Jakobsson's algorithm runs emitting $n/2 = 2^k$ hash elements
that the remaining pebbles contain hash elements of hash chain positions

\begin{eqnarray*}
2^1, 2^2, 2^3, \ldots, 2^k.
\end{eqnarray*}

\noindent
{\bf Inductive Step:} Consider $n/2$ hash element exposures for a hash chain with 
$n=2^{k+1}$ total elements where $k \geq 2$.  This step will show that the pebbles will
be placed at positions equivalent to $2^1, 2^2, \ldots, 2^{k}$.

Again, by Fact~\ref{F5} only one pebble, the pebble associated with the $2^{k}$ position will be discarded.
This leaves $k-1$ pebbles remaining.  One of these pebbles is 
associated with the root and will not change.

By Fact \ref{F4} for all pebbles $p[{\bf j} \overset{k}{\rightarrow} 1]$, it must be that
\begin{eqnarray*}
p[{\bf j} \overset{k}{\rightarrow} 1].{\bf Position} & = & p[{\bf j} \overset{k}{\rightarrow} 1].{\bf Destination}.
\end{eqnarray*}

\noindent
For all remaining pebbles after $n/2$ hash exposures,
\begin{eqnarray*}
p[j].{\bf Position} & = & P(j,q), \\
p[j].{\bf Destination} & = & D(j,q).
\end{eqnarray*}

By Corollary~\ref{C1} we know that for all remaining pebbles 
$p[j].{\bf Position} = p[j].{\bf Destination}$ when $n/2$ elements are output and thus all 
pebbles are inactive.

Further, since we know that $p[j].{\bf Position} = p[j].{\bf Destination}$ the remaining 
pebble positions are defined by the destination function $D(j,m)$ as defined earlier where
$m$ is the number of times the pebble at original position $j$ is moved back and
\begin{eqnarray*}
D(j,m) = 2^j + m \cdot 2^{j+1}.
\end{eqnarray*}

By Fact~\ref{F7} $m = 2^{k-j-1}$ for each pebble at initial position $j$ over the first
$n/2$ hash element exposures.

Therefore, 
\begin{eqnarray*}
D(j,2^{k-j-1}) & = & 2^j + 2^{k-j-1} \cdot 2^{j+1}\\
D(j,2^{k-j-1}) & = & 2^j + 2^k
\end{eqnarray*}

\noindent
given in terms of base $n=2^{k+1}$ total elements.  To put this result in terms of our new 
hash chain $n=2^k$ we simply apply the $R_{k}(l_{k+1})$ function to get  
\begin{eqnarray*}
R_{k}(2^j + 2^k)= 2^j + 2^k - 2^k = 2^j,
\end{eqnarray*}

\noindent
where $j = 1, \cdots, k$.

Thus, by the inductive hypothesis it must be that when Jakobsson's algorithm runs emitting $n/2 = 2^{k}$ hash elements
the remaining pebbles contain hash elements of hash-chain positions 
$2^1, 2^2, 2^3, \ldots, 2^k$, thus completing the proof.
\end{proof}

\end{document}